\documentclass[11pt]{article}
\pdfoutput=1

\usepackage{array}
\usepackage{epsfig}
\usepackage{amssymb}
\usepackage{slashed}
\usepackage{amsmath}
\usepackage{graphicx}
\usepackage{cite}
\usepackage{ifpdf}

\newcommand{\beq}{\begin{equation}}
\newcommand{\eeq}{\end{equation}}
\newcommand{\be}{\begin{equation}}
\newcommand{\ee}{\end{equation}}
\newcommand{\ba}{\begin{array}}
\newcommand{\ea}{\end{array}}
\newcommand{\beqa}{\begin{eqnarray}}
\newcommand{\eeqa}{\end{eqnarray}}
\newcommand{\bea}{\begin{eqnarray}}
\newcommand{\eea}{\end{eqnarray}}
\newcommand{\beqn}{\begin{eqnarray}}
\newcommand{\eeqn}{\end{eqnarray}}

\newcommand{\D}{\Delta}

\newcommand{\eps}{\epsilon}
\newcommand{\nn}{\nonumber}

\newcommand{\re}{{\rm Re}}
\newcommand{\im}{{\rm Im}}

\newcommand{\bra}[1]{\ensuremath{\langle #1 |}}
\newcommand{\ket}[1]{\ensuremath{| #1 \rangle }}
\newcommand{\cA}{{\cal A}}
\newcommand{\cO}{{\cal O}}

\newcommand{\cL}{{\cal L}}

\newcommand{\cH}{{\cal H}}
\newcommand{\cT}{{\cal T}}

\newcommand{\epsp}{\epsilon^\prime}
\newcommand{\epsb}{\bar\epsilon}
\newcommand{\epsK}{\epsilon_K}

\begin{document}

\begin{flushright}
  TUM-HEP-751/10
\end{flushright}

\medskip

\begin{center}
{\Large \bf \boldmath On $\epsK$ beyond lowest order in the Operator Product Expansion}\\[0.8 cm]
{\large Andrzej~J.~Buras$^{a,b}$, Diego Guadagnoli$^c$, and Gino Isidori$^{b,d}$} \\[0.5 cm]
\small
$^a${\em Physik-Department, Technische Universit\"at M\"unchen, James-Franck-Stra{\ss}e,
\\D-85748 Garching, Germany} \\[0.1cm]
$^b${\em TUM Institute for Advanced Study, Technische~Universit\"at~M\"unchen, Arcisstra{\ss}e 21,
\\D-80333 M\"unchen, Germany} \\[0.1cm]
$^c${\em Excellence Cluster Universe, Technische Universit\"at M\"unchen, Boltzmannstra{\ss}e 2,
\\D-85748 Garching, Germany}\\[0.1cm]
$^d${\em INFN, Laboratori Nazionali di Frascati, I-00044 Frascati, Italy} \\[0.8 cm]
\end{center}

\abstract{%
\noindent We analyse the structure of long distance (LD) contributions to the 
CP-violating parameter $\epsilon_K$, that generally affect both the {\it absorptive} 
($\Gamma_{12}$) and the {\it dispersive} ($M_{12}$) parts of the $K^0-\bar K^0$ mixing amplitude. 
We point out that, in a consistent framework, in addition to 
LD contributions to $\im \Gamma_{12}$, estimated recently by two of us, also LD 
contributions to $\im M_{12}$ have to be taken into account. Estimating the latter 
contributions the impact of LD effects on $\epsilon_K$ is significantly
reduced (from $-6.0\%$ to $-3.6\%$). 
The overall effect of LD corrections and of the superweak 
phase being different from $45^\circ$ is summarised by the multiplicative factor 
$\kappa_\epsilon =0.94\pm 0.02$.
}

\section{Introduction}

Some of the most important tests of the Standard Model (SM) are offered by  
CP-violating observables, that in this model are supposed to originate from a
single CP-odd phase in the CKM matrix~\cite{Cabibbo:1963yz}.
In particular, the crucial test is the
hierarchy of CP-violating effects in $B_d$, $B_s$ and $K$ systems predicted by
this model. Indeed the most prominent CP-violating observables in these
three systems, $S_{\psi K_S}$, $S_{\psi\phi}$ and $\eps_K$, predicted by 
the SM, differ by orders of magnitude
\be
 S_{\psi K_S}\approx 2/3,\qquad  S_{\psi\phi}\approx 4\times 10^{-2}, \qquad 
|\eps_K|\approx 2\times 10^{-3}~.
\ee

Extensive analyses of the Unitarity Triangle have shown a spectacular
consistency of the data for $S_{\psi K_S}$ and $\eps_K$, within the 
parametric and theoretical uncertainties in $\eps_K$, that until recently
were rather sizable. The size of $S_{\psi\phi}$ measured by CDF~\cite{Aaltonen:2007he} 
and D\O{}~\cite{:2008fj}  appears
to be by one order of magnitude larger than predicted by the SM, but the 
large experimental errors preclude any definitive conclusions.

Recently the consistency of the measured values for $S_{\psi K_S}$ and 
$\eps_K$ within the SM has been challenged in \cite{Lunghi:2008aa,Buras:2008nn}
due to two facts:
\begin{itemize}
\item  The improved value of the relevant hadronic parameter 
$\hat B_K$ from unquenched lattice QCD that enters the evaluation of
$\eps_K$. This parameter is now not only known with an accuracy 
of $4\%$ \cite{Antonio:2007pb,Aubin:2009jh} but turns out to be significantly lower than previously found 
in lattice calculations, suppressing by $10\%$ the previous estimates of $\eps_K$.

\item A more careful look at $\eps_K$, that identified an 
additional suppression of $|\eps_K|$, summarised by  
a multiplicative factor $\kappa_\eps=0.92\pm0.02$~\cite{Buras:2008nn} 
to the previously adopted formula for $\epsilon_K$.
\end{itemize}

In view of these two suppressions, as demonstrated in~\cite{Buras:2008nn},
the size of CP violation measured in $B_d\to\psi K_S$ might be insufficient
to describe $\eps_K$ within the SM. Clarifying this new tension 
is important as the $S_{\psi K_S}-\eps_K$ correlation in the SM is 
presently the most important direct relation between CP violation in the 
$B_d$ and $K$ systems that can be tested experimentally. 

The correction calculated in \cite{Buras:2008nn} originates from two 
factors: i) the difference of the superweak phase $\phi_\eps$ from $45^\circ$, 
and ii) the long-distance contribution to $\eps_K$
arising from the imaginary part of the {\it absorptive}  amplitude 
of the $K^0-\bar K^0$ mixing, $\Gamma_{12}$. The latter effect has been 
estimated with the help of the $\Delta I=1/2$ dominance in $K\to\pi\pi$ 
decays and the experimental value for $\eps^\prime/\eps$.

In the present paper we point out that at the same level of accuracy
other effects should be considered, in particular the long distance 
contributions to the imaginary part of the {\it dispersive}  amplitude $M_{12}$.
While this topic has been the subject of intensive discussions in the 
mid 1980's, it is important to have a fresh look at this issue in view
of the decrease of the error in $\hat B_K$ and of the theoretical advances
during the last twenty five years.

Our paper is organized as follows. In Section \ref{sec:notation} we present 
general formulae { from which the different contributions to $\eps_K$ can be 
clearly identified.} In Section \ref{sec:OPE} we discuss $\eps_K$ using the Operator 
Product Expansion (OPE). This allows us to identify the most important, { still missing,}
long-distance contributions to $\im M_{12}$. In Section \ref{sec:CHPT} we estimate the size 
of these contributions in the framework of Chiral Perturbation Theory (CHPT), and briefly 
compare our findings with previous literature. We conclude in Section \ref{sec:conclusions}.

\section{Notation and general formulae}\label{sec:notation}

Indirect CP violation originates in the weak phase difference between the (off-diagonal elements 
of the) Hermitian matrices $M$ and $\Gamma$ which control the time evolution of a neutral 
meson system. For the $K^0 - \bar{K}^0$ system one has
\beq
i \frac{d}{dt} \left(\ba{c} \ket{K^0(t)} \\ \ket{\bar K^0(t)} \ea \right)
= \left( M - i\, \frac{\Gamma}{2} \right) 
\left(\ba{c} \ket{K^0(t)} \\ \ket{ \bar K^0(t)} \ea \right)~.
\label{mgmat}
\eeq
Defining the eigenvectors
\beqn
\label{eigenvectors}
\ket{K_{S(L)}} =\frac{1}{\sqrt{2(1+|\epsb|^2)}}\left[ (1 + \epsb)\ket{K^0} \mp (1 - \epsb) \ket{\bar K^0} \right]~,
\eeqn
the following phase-convention-independent relation 
holds:
\be
\frac{ \re(\epsb) }{ 1 + |\epsb|^2} = 
\frac{\im(\Gamma_{12} M_{12}^*) }{ 4 |M_{12}|^2 +|\Gamma_{12}|^2}
\left[ 1 +\cO\left(\im\left(\frac{\Gamma_{12}}{M_{12}}\right)\right)\right]~.
\label{eq:eps0}
\ee
This represents indeed the indirect CP-violating parameter measured 
from the semileptonic charge asymmetries~\cite{AlaviHarati:2002bb}
or the Bell-Steinberger relation~\cite{Ambrosino:2006ek}.
The experimental smallness of $\re(\epsb)$ makes the expansion 
to first non-trivial order in the weak phases an excellent 
approximation. At this level of accuracy we can identify 
$\re(\epsb)$ with the real part of the complex 
quantity $\epsK$, defined in terms of the $K \to 2 \pi$ amplitudes,
\be
\epsK  = \frac{2 \eta_{+-} +\eta_{00}}{3}~, 
\qquad \eta_{ij}= \frac{\cA(K_L\to \pi^i\pi^j)}{\cA(K_S\to \pi^i\pi^j)}~.
\ee
The two parameters are indeed related by $\epsK  = \epsb + i \xi$,
where $\xi$ is the weak phase of the $K^0 \to (2\pi)_{I=0}$ amplitude, namely
\be
\xi  = \frac{\im A_0}{\re A_0}~, ~~~A_0 \equiv \cA(K^0 \to (2\pi)_{I=0})~.
\ee

Expanding to first non-trivial order in the weak phases we have
\bea
\label{eq:DmDGa}
\Delta m_K &=& m_L - m_S =  2 \re(M_{12})~, \nn \\
\Delta \Gamma &=& \Gamma_S - \Gamma_L  =  - 2 \re(\Gamma_{12})~.
\eea
Introducing also the so-called superweak phase,
$\phi_\epsilon = \arctan\left( 2 \Delta m_K/\Delta \Gamma \right)$, 
the expression for $\re(\epsb)$ becomes
\be
\re(\epsK)  = \re(\epsb) = \cos\phi_\epsilon\sin\phi_\epsilon 
\left[ \frac{\im M_{12}}{ 2 \re M_{12} } - \frac{\im \Gamma_{12} }{ 
2 \re \Gamma_{12} } \right]~.
\label{eq:eps1}
\ee
A further simplification arises by the observation that the $\ket{(2\pi)_{I=0}}$
final state largely saturates the neutral kaon decay widths.
Since
\be
\Gamma_{21} = \Gamma_{12}^* = \sum_{f} \cA(K^0\to f) \cA(\bar K^0 \to f)^* ~,
\ee
the $\ket{(2\pi)_{I=0}}$ dominance in the sum over final states implies
\be
\frac{\im \Gamma_{12} }{ \re\Gamma_{12} } \approx - 2 \frac{\im A_0}{\re A_0}
= - 2 \xi~.
\label{eq:xi}
\ee
Expressing $\re M_{12}$ in terms of $\Delta m_K$ and using Eq.~(\ref{eq:xi})
we arrive at
\be
\re(\epsK)  = \cos\phi_\epsilon\sin\phi_\epsilon 
\left[ \frac{\im M_{12}}{\Delta m_K } +\xi  \right]~,
\label{eq:eps2}
\ee
which is consistent with
\be
\epsK  = e^{i\phi_\epsilon} \sin\phi_\epsilon 
\left[ \frac{\im M_{12}}{\Delta m_K } +\xi  \right]~.
\label{eq:eps2K}
\ee
The equation above allows us to calculate $\epsilon_K$ by taking $\phi_\epsilon$ 
and $\Delta m_K$ from experiment and calculating $\im M_{12}$ and $\xi$ in a given 
model, in particular the SM. In Ref.~\cite{Buras:2008nn} only {\it short distance} 
contributions to $\im M_{12}$, represented by the well known box diagrams, 
have been included, while $\xi$ has been calculated by relating it to the 
ratio $\epsilon^\prime/\epsilon$ and taking the latter from experiment. 
As we will discuss in the following, this approach is not fully consistent:
in this way $\im \Gamma_{12}$ and  $\im M_{12}$ are evaluated at a 
different order in the OPE. In particular, {\it long distance} contributions
to $\im M_{12}$, which are of the same order of 
$\im \Gamma_{12}$ (the latter giving rise to the $\xi$ term in Eq.~(\ref{eq:eps2K})), 
are missing.

\section{\boldmath Decomposition of Re($\epsK$) using the OPE} \label{sec:OPE}

As shown in Eq.~(\ref{eq:eps1}) the evaluation of $\eps_K$ 
requires the knowledge of the weak phases of both $M_{12}$ and $\Gamma_{12}$. 
In this respect, we should emphasize that $\im M_{12}$ and $\im \Gamma_{12}$ are both 
generated at $\cO(G_F^2)$. Since $\re M_{12}$ and $\re \Gamma_{12}$ are 
very similar in size ($\phi_\epsilon \approx 43.5^\circ$), we 
should consistently evaluate $\im M_{12}$ and $\im \Gamma_{12}$ 
at the same order in the OPE.

\begin{figure}[t]
\begin{center}
\includegraphics[width=0.35\textwidth]{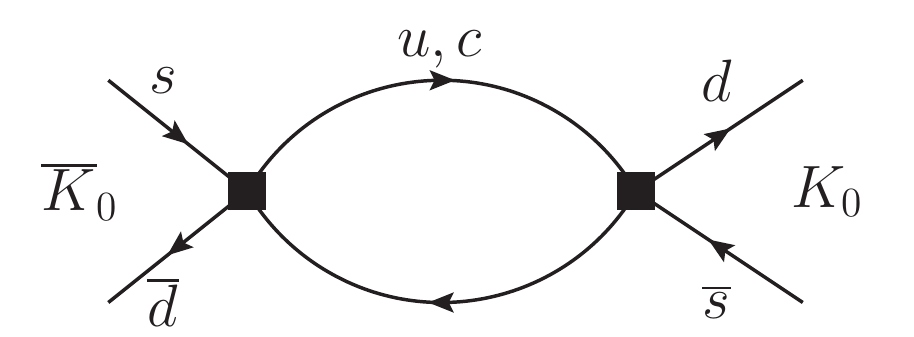}
\hspace{0.3cm}
\includegraphics[width=0.35\textwidth]{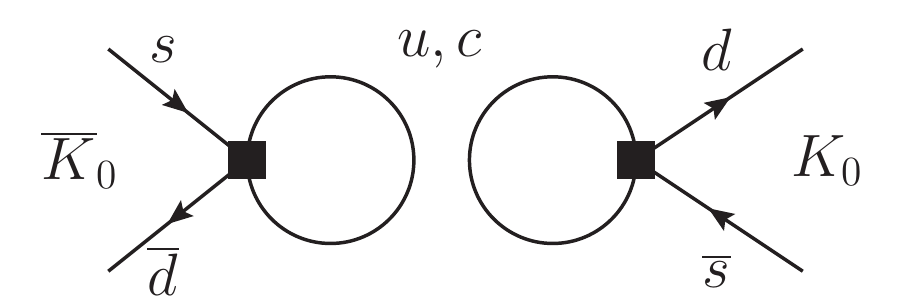}
\end{center}
\caption{\label{fig:quark} Contractions of the leading $|\Delta S|=1$ 
four-quark effective operators contributing to $M_{12}$ at $\cO(G_F^2)$.}
\end{figure}

The relevant effective Hamiltonians are $\cH_{\Delta S=2}$ (contributing 
to $\im M_{12}$ only) and $\cH_{\Delta S=1}$ (contributing to both 
$\im M_{12}$  and $\im \Gamma_{12}$).
The leading term in the OPE is the short-distance contribution 
to $\im M_{12}$,
\be
\label{eq:imM126}
\im M_{12}^{(6)} \equiv \im M_{12}^{SD} = \frac{1}{2 m_K} \im \left(
\bra{\bar K^0} \cH^{(6)}_{\Delta S=2} \ket{K^0} \right)^*
\ee
where 
\be
\cH^{(6)}_{\Delta S=2} =  \frac{G_F^2 m_W^2}{16\pi^2} \times F_0 \times 
Q^{(6)}~,
\qquad 
Q^{(6)} =(\bar s d)_{V-A} (\bar s d)_{V-A}~, 
\label{eq:db2}
\ee
is the dimension-six $\Delta S=2$ effective Hamiltonian.
The operator $Q^{(6)}$ does not mix with other operators
and the imaginary part of its Wilson coefficient is dominated by terms proportional 
to the top-quark Yukawa coupling.\footnote{~The 
explicit expression of the coefficient function $F_0$, depending on 
quark masses and CKM elements, 
can be found in~\cite{Buchalla:1995vs}.}
At this order in the OPE one is neglecting 
terms generated by two insertions of
$\Delta S=1$ operators (see Fig.~\ref{fig:quark}) which cannot be
absorbed into the coefficient of $Q^{(6)}$.
For consistency, this implies one should
set $\im \Gamma_{12}$ to zero, since $\im \Gamma_{12}$ is 
the absorptive part of the diagrams in Fig.~\ref{fig:quark}.
In other words, 
the leading order result is obtained 
with the following  substitutions in Eq.~(\ref{eq:eps2}):
\be
\im  M_{12} \to  \im M_{12}^{(6)} = \im M_{12}^{SD} \qquad {\rm and} \qquad \xi \to 0~.
\ee
Going one step forward requires taking into account:
\begin{enumerate}
\item non-local contributions to both $\im M_{12}$ and 
$\im \Gamma_{12}$ generated by the $\cO(G_F)$
dimension-six $\Delta S=1$ operators,
\item local contributions to $\im M_{12}$ generated by 
dimension-eight $\Delta S=2$ operators of $\cO(G_F^2)$.
\end{enumerate}
The structure of the subleading terms in $\im M_{12}$
is very similar to the $\cO(G_F^2)$ long-distance 
contributions to $K\to\pi\nu\bar\nu$, discussed in Ref.~\cite{Isidori:2005xm}.
The relevant effective Hamiltonian
changes substantially if we choose a renormalization 
scale above or below the charm mass.
Keeping the charm as explicit degree of freedom,
dimension-eight operators are safely negligible 
and the key quantity to evaluate is 
\be
\cT_{12}  = -i \int d^4 x \bra{K^0} 
T\left[ \cH^{(u,c)}_{|\Delta S|=1} (x)  \cH^{(u,c)}_{|\Delta S|=1} (0) \right]
\ket{\bar K^0}~,
\label{eq:Tprodc}
\ee
where the superscript in $\cH^{(u,c)}_{\Delta S=1}$
denotes that the we have two
dynamical up-type quarks. The absorptive part of 
$\cT_{12}$ contributes to $\Gamma_{12}$, while the 
dispersive part contributes to  $M_{12}$. In the latter case 
the leading term in the expansion in local operators
should be subtracted, being already included in $\im M_{12}^{(6)}$.
In principle, extracting the subleading contribution to $\im M_{12}$ 
directly from Eq.~(\ref{eq:Tprodc}) is the best strategy: the result would 
be automatically scale independent. However, in practice this is 
far from being trivial also on the lattice, given the disconnected 
diagrams in Fig.~\ref{fig:quark}.

Following a purely analytical approach, we can 
integrate out the charm and renormalize $\cH_{\Delta S=1}$ below the charm mass.
This allows to identify $\xi$ with the weak
phase of the $A_0$ amplitude, that, as mentioned, has already been estimated
in Ref.~\cite{Buras:2008nn} (see also \cite{Buras:2003zz}). On the other hand,
$\im  M_{12}$ assumes the form
\be
\im  M_{12} = \im M_{12}^{SD} + \im M_{12}^{LD}~, \qquad 
  \im M_{12}^{LD} = \im M_{12}^{\rm non-local}  
+ \im M_{12}^{(8)}~,
\ee
where $\im M_{12}^{\rm non-local}$ and $\im M_{12}^{(8)}$ are not 
separately scale independent. The structure of the dimension-eight operators obtained 
integrating out the charm, and an estimate of their impact 
on $\epsK$, has been presented in Ref.~\cite{Cata:2003mn}.
According to this estimate, $\im M_{12}^{(8)}$ is less than 
 $1\%$ of the leading term.  

The smallness of $\im M_{12}^{(8)}$ can be understood by the 
following dimensional argument. First, it should be noted 
that the CKM suppression of the dimension-eight operators 
is $(V_{cs}^*V_{cd})^2$, namely the same CKM factor
of the genuine charm contribution in $\cH^{(6)}_{\Delta S=2}$. 
Second, even if we are 
not able to precisely evaluate the hadronic matrix elements 
of the dimension-eight operators, we expect 
\be
\label{eq:Qi8naive}
\bra{\bar K^0} Q_i^{(8)} \ket{K^0} = \cO(1) \times m^2_K \times  
\bra{\bar K^0} Q^{(6)} \ket{K^0}~.
\ee
According to this scaling, the contribution of 
$\im M_{12}^{(8)}$ is an $\cO(m_K^2/m_c^2 \approx 15\%)$ correction
of the charm contribution (charm-charm loops) 
to $\im M_{12}^{(6)}$, which itself is an $\cO(15\%)$ correction 
of the total dimension-six contribution.
We are thus left with an overall $\cO(2\%)$ naive suppression of 
 $\im M_{12}^{(8)}$
with respect to  $\im M_{12}^{(6)}$. According to the explicit evaluation 
in Ref.~\cite{Cata:2003mn}, the actual numerical impact is
even smaller.

The only potentially large long-distance contribution to $\im M_{12}$ is 
the contribution of the non-local terms enhanced by the $\Delta I=1/2$ rule.
For this purpose, we observe that if we had a single weak 
operator in $\cH_{\Delta S=1}$, this would generate the same 
weak phase to both $\im M_{12}^{LD}$ and  $\im \Gamma_{12}$.
As we discuss in more detail in the next section,  
this is what happens to lowest order in CHPT, 
where the  $\Delta I=1/2$ part $\cH_{\Delta S=1}$ has 
only one operator, with effective coupling $G_8$.
Decomposing  $\im M_{12}^{LD}$ as a leading term 
proportional to $G_8^2$, and a subleading term with different
effective coupling 
\be
\im M_{12}^{LD} = \left.\im M_{12}^{LD}\right|_{G_8^2} +   \left. \im M_{12}^{LD}\right|_{{\rm non}-G^2_8}~,
\ee
we can write 
\be
\left.\im M_{12}^{LD}\right|_{G_8^2} = \left.\re M_{12}^{LD}\right|_{G_8^2} 
\times  \frac{\im [(G_8^*)^2] }{ \re [(G_8^*)^2] }~,
\ee
and identify the weak phase of $G_8$ with $\xi$. As a result,
\be
\left.\im M_{12}^{LD}\right|_{G_8^2} \approx  \left.\re M_{12}^{LD}\right|_{G_8^2} 
\times (-2\xi) \approx  - \xi \times \left(
\Delta m^{LD}_K|_{G_8^2} \right)~.
\label{eq:main}
\ee
This allow us to re-write Eq.~(\ref{eq:eps2}) as follows
\be
\re(\epsK)  = \cos\phi_\epsilon\sin\phi_\epsilon 
\left[ \frac{\im M^{(6)}_{12}}{\Delta m_K } +\xi  
\left( 1 -\frac{\Delta m^{LD}_K|_{G_8^2}}{\Delta m_K} \right) 
+  \delta_{ \im M_{12} } \right]~,
\label{eq:eps_new}
\ee
where $\delta_{ \im M_{12} }$ encodes the subleading terms in $\im M_{12}^{LD}|_{{\rm non}-G^2_8}$
(including also  $\im M_{12}^{(8)}$).
Note that, in the limit where the contribution of $G_8$ saturates $\Delta m_K$, the contribution 
of $\xi$ would be absent. 
This is exactly what we should expect, since in this limit $M_{12}$ and $\Gamma_{12}$
would have the same weak phase but for the short-distance contribution 
to $\im M_{12}$.

\section{Estimate of long-distance effects in CHPT} \label{sec:CHPT}

A convenient framework for estimating the long-distance 
contribution to $M_{12}$ is provided by
Chiral Perturbation Theory (CHPT).
In this framework 
$\pi$, $K$ and $\eta$ fields are identified 
with the would-be Goldstone bosons arising 
from  the $SU(3)_L\times SU(3)_R \to SU(3)_{L+R}$
symmetry breaking of the QCD action in the 
limit of vanishing light quark masses. Low-energy amplitudes
involving these mesons, expanded in powers of their 
masses and momenta, are evaluated by means of an effective 
Lagrangian written in terms of 
the pseudo-Goldstone boson fields.

The lowest-order effective Lagrangian describing 
non-leptonic $\Delta S=1$ decays has only two operators,
transforming as  $(8_L,1_R)$  and $(27_L,1_R)$ 
under the  $SU(3)_L\times SU(3)_R$
chiral group. Moreover, only the $(8_L,1_R)$
operator has a phenomenologically large coefficient,
being responsible for the enhancement of 
$\Delta I=1/2$ amplitudes. As a result,
the only term in the effective Lagrangian 
relevant to our calculation is
\be
\cL^{(2)}_{|\Delta S| = 1} = F^4  G_8  
\left( \partial^\mu U^\dagger \partial_\mu U \right)_{23} +{\rm h.c.} ~,
\ee
where, as usual, we define 
\be
U =\exp(i \sqrt{2} \Phi/F)~, \qquad \Phi=\left[
\begin{array}{ccc}
\frac{\pi^0}{\sqrt{2}}+\frac{\eta}{\sqrt{6}} & \pi^+ & K^+ \\
\pi^- & -\frac{\pi^0}{\sqrt{2}}+\frac{\eta}{\sqrt{6}} & K^0 \\
K^- & \bar K^0 & -\frac{2\eta}{\sqrt{6}} \\
\end{array}\right]~,
\ee
and $F$ can be identified with the pion decay constant
($F\approx 92\, {\rm MeV}$). The effective coupling $G_8$ 
can be determined by $K\to 2\pi$ amplitudes. Neglecting 
the $(27_L,1_R)$ operator and evaluating the $K\to 2\pi$ 
amplitudes at tree level leads to
\be
A_0 = \cA(K^0 \to (2\pi)_{I=0}) = \sqrt{2}  F G_8 (m_K^2 -m_\pi^2)~,
\ee
which implies $|G_8|\approx 9\times 10^{-6}~({\rm GeV})^{-2}$.
As far as the weak phase of $G_8$ is concerned, at this 
level of accuracy we have $\im(G_8)/\re(G_8) =\xi$.

\begin{figure}[t]
\begin{center}
\raisebox{0.8cm}{\includegraphics[width=0.30\textwidth]{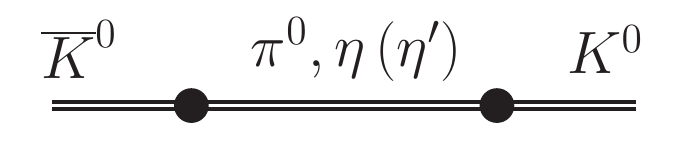}}
\hspace{1cm}
\includegraphics[width=0.30\textwidth]{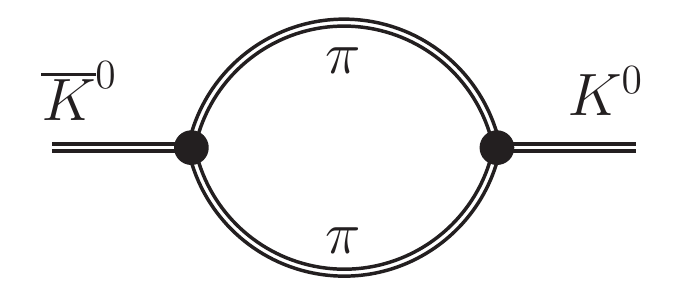}
\end{center}
\caption{\label{fig:chpt} Tree-level and one-loop diagrams 
contributing to  $\bar K^0$--$K^0$ mixing 
in CHPT.}
\end{figure}

In principle $\cL^{(2)}_{|\Delta S| = 1}$ could contribute to 
$M_{12}$ already at $\cO(p^2)$, via the 
tree-level diagram in Fig.~\ref{fig:chpt} (left). 
However, considering the  $\cO(p^2)$ relation 
among $\pi^0$, $\eta$ and kaon masses
(i.e.~the Gell-Mann--Okubo mass formula), this 
contribution vanishes~\cite{Donoghue:1983hi}.
As a result, the first non-vanishing contribution 
to $M_{12}$ generated by  $\cL^{(2)}_{|\Delta S| = 1}$
arises only at $\cO(p^4)$.

At $\cO(p^4)$ we should evaluate loop amplitudes 
with two insertions of $\cL^{(2)}_{|\Delta S| = 1}$
and tree-level diagrams with the insertion of 
appropriate $\cO(p^4)$ counterterms. 
Among all these $\cO(p^4)$ contributions, the only model-independent, 
and presumably dominant, contribution to $M_{12}$ 
is the non-analytic one generated by the pion-loop 
amplitude in Fig.~\ref{fig:chpt} (right),
\bea
&& T_{12}^{(\pi \pi)} = 
\cA^{(\pi \pi)}(\bar K^0 \to K^0)  = - \frac{3}{16 \pi^2} F^2 (G_8^*)^2 (m_K^2 - m_\pi^2)^2 \times
\nn \\
&& \qquad\qquad  \times \left[ \sqrt{1 - 4r_\pi^2 } \left( \log \frac{1 + \sqrt{1 - 4r_\pi^2 } }{1 
- \sqrt{1 - 4r_\pi^2 }} -  i \pi \right)  + \log\left(\frac{m_\pi^2}{\mu^2}\right)  \right]~, 
\label{eq:App}
\eea
with $r_\pi^2 = m_\pi^2/m_K^2$ and where we have absorbed all finite (mass-independent) terms in the definition 
of the renormalization scale $\mu$. This is the only contribution which has an absorptive part.
As a consequence, its weak phase can be unambiguously related to the weak phase 
of the $K^0 \to (2 \pi)_{I=0}$ amplitude to all orders in the chiral expansion. In addition, 
it is the only contribution that survives in the limit of $SU(2)_L \times SU(2)_R$ CHPT, which 
is known to represent a good approximation of the full $\cO(p^4)$ amplitude in several $K$-decay
observables where contributions from counterterms are fully under control 
(see e.g.~Ref.~\cite{Colangelo:2000zw}).

A CHPT calculation of $M_{12}$ complete to $\cO(p^4)$ would require consideration of loops 
involving kaons and $\eta$'s, as well as $\cO(p^4)$ local counterterms. However, 
all these additional pieces are not associated with any
physical cut. As such, they can effectively be treated as a local term
whose overall weak phase cannot be related to the phase of the $K^0 \to (2 \pi)_{I=0}$ amplitude.\footnote{For 
a recent, elucidating discussion about the role of kaon loops in CHPT, see~\cite{Donoghue:SU3CHPT}.} 
On account of the above considerations,\footnote{The authors 
warmly acknowledge Jean-Marc G\'erard for triggering a discussion on this point.}
we refrain from a full $\cO(p^4)$ CHPT calculation,
and we focus on the pion-loop non-analytic contribution only.
Using the relation $T_{12}^{(\pi \pi)} = 2 m_K M_{12}^{\rm (\pi \pi)}(\mu)$, 
the result in Eq. (\ref{eq:App}) implies
\be
M_{12}^{\rm (\pi \pi)}(\mu)
=  - \frac{3}{64 \pi^2 m_K} (A_0^*)^2 \left[ \log\left(\frac{m_K^2}{\mu^2}\right) 
 + \cO\left(\frac{m_\pi^2}{m_K^2}\right) \right]~.
\label{eq:Mpp}
\ee
The absorptive part in Eq.~(\ref{eq:App}) is nothing but the leading
$\ket{(2\pi)_{I=0}}$ contribution to $\Gamma_{12}$, which gives rise to the 
relation (\ref{eq:xi}). The dispersive part
is the dominant contribution to $M_{12}$ in the leading-log approximation.
The close link of these two terms is a further confirmation that we 
cannot neglect the long-distance contribution to  $\im M_{12}$ 
if we want to keep track of all the $\cO(\xi)$ terms in $\epsK$.

Using the result in Eq.~(\ref{eq:Mpp}) we can estimate the contribution to  $\im M_{12}$ 
proportional to $G_8$, which enters in the phenomenological
formula for $\re(\epsK)$ in Eq.~(\ref{eq:eps_new}). 
Setting $\mu =800$~MeV and varying it in the interval $0.6 \div 1$~GeV leads to 
\beqn
\frac{\Delta m^{LD}_K|_{G_8^2}}{\D m_K^{\rm exp}} = \frac{2 \re M_{12}^{\rm (\pi \pi)} }{\D m_K^{\rm exp}} = 0.4 \pm 0.2~.
\label{eq:chpt_main}
\eeqn
Note that the result has a well-defined sign since $G_8$ (or $A_0$) appears
squared in $M_{12}^{\rm (\pi \pi)}$. Using this result in  Eq.~(\ref{eq:eps_new})
we find a suppression of the $\xi$ term relative to the estimate in 
\cite{Buras:2008nn}, where only the LD contribution to $\im \Gamma_{12}$ has been taken
into account.

\medskip

Since our estimate of $\Delta m^{LD}_K|_{G_8^2}$ is not the result of a complete 
calculation at fixed order in the chiral expansion, it is worthwhile to cross-check
it using a different argument. For this purpose, we note that
the only relevant contribution to $M_{12}$, beside the two-pion intermediate state, 
is expected to arise from the tree-level  $\eta^\prime$ exchange 
(Fig.~\ref{fig:chpt} left)~\cite{Gerard:2005yk}. We can thus decompose $M_{12}$
as follows:
\be
M_{12} \approx M^{SD}_{12} +  M^{LD}_{12}|_{\pi\pi} +  M^{LD}_{12}|_{\eta^\prime}~. 
\ee
According to this decomposition 
it is clear that, as far as long-distance contributions are concerned, we can 
trade the evaluation of $ M^{LD}_{12}|_{\pi\pi}$ for that of $M^{LD}_{12}|_{\eta^\prime}$.
An estimate of the  $\eta^\prime$ contribution to $M_{12}$
goes beyond pure CHPT, where it can be considered as a free parameter
(the leading contribution to the $\cO(p^4)$ local terms).
However, its impact can be estimated in the 
large $N_c$ limit, extending the underlying symmetry 
from  $SU(3)_L\times SU(3)_R$ to $U(3)_L\times U(3)_R$. 
Within this framework the operator basis 
must be extended and we cannot directly relate the phase of 
the $\eta^\prime$ exchange amplitude to the phase of $G_8$.
According to the recent analysis in Ref.~\cite{Gerard:2005yk}, 
the $\eta^\prime$ amplitude gives a negative contribution to 
$\D m_K$:
\be
 2 \re  M^{LD}_{12}|_{\eta^\prime} = \Delta m^{LD}_K|_{\eta^\prime} \approx -0.3 \D m_K^{\rm exp}~.  
\label{eq:DMKeta}
\ee
Most important for our analysis, this contribution 
is found to be induced at the quark level by  the 
operator $(\bar s d)_{V-A} \times (\bar u u)_{V-A}$ only~\cite{Gerard:2005yk}.
This implies that the $\eta^\prime$ exchange has a vanishing weak phase 
in the standard CKM phase convention:
\be
\im M_{12}^{LD}|_{\eta^\prime} =0~.
\ee
Using this result in Eq.~(\ref{eq:eps2}), and using the relation (\ref{eq:main}) 
for the $\pi\pi$ contribution, we get 
\be
\re(\epsK)  = \cos\phi_\epsilon\sin\phi_\epsilon 
\left[ \frac{\im M^{(6)}_{12}}{\Delta m^{\rm exp}_K } +\xi \,
\frac{\Delta m_K^{SD}+ \Delta m^{LD}_K|_{\eta^\prime}}{\Delta m_K^{\rm exp} }  \right]~,
\label{eq:eps_eta}
\ee
where the $G_8$ term (i.e.~the $\pi\pi$ contribution), is manifestly absent.
Denoting as $\rho$ the coefficient of the $\xi$ term in Eq.~(\ref{eq:eps_eta}), 
and combining Eq.~(\ref{eq:DMKeta}) with the NLO short-distance estimate of $\re M_{12}$,
namely $\Delta m_K^{SD} = (0.7\pm 0.1) \D m_K^{\rm exp}$~\cite{HerrlichNierste,Buchalla:1995vs}, 
we get $\rho = 0.4\pm 0.1$. This result is well consistent with the value 
$\rho= 0.6\pm 0.2$ obtained from Eq.~(\ref{eq:eps_new}) with the direct 
evaluation of the $\pi\pi$ contribution in Eq.~(\ref{eq:chpt_main}).

\medskip

We rate the direct evaluation of the $\pi\pi$ loop as the most
reliable estimate of $\rho$. As a consequence, our final
phenomenological expression for $\epsK$ is
\be
\epsK  = \sin\phi_\epsilon e^{i\phi_\eps}
\left[ \frac{\im M^{(6)}_{12}}{\Delta m_K } + \rho \, \xi  \right] \qquad
{\rm with} \qquad  \rho = 0.6 \pm 0.3~,
\label{eq:eps_phen}
\ee
where we have conservatively increased by $50\%$ the error in  Eq.~(\ref{eq:chpt_main}) 
to take into account the sub-leading contributions of $\im M_{12}^{LD}|_{{\rm non}-G^2_8}$.
For $\rho=1$ our result reduces to the one in \cite{Buras:2008nn}. The contribution 
calculated in this paper, resulting in  $\rho < 1$,
completes the estimate of the terms of $\cO(\xi)$ in $\eps_K$.
 
Following the notation of Ref.~\cite{Buras:2008nn}, we summarise the corrections
to $\eps_K$ due to LD effects and $\phi_\eps\not=45^\circ$, 
via the introduction of the phenomenological 
factor $\kappa_\eps$, defined by 
\be
\epsK  = \kappa_\eps \frac{ e^{i\phi_\eps} }{\sqrt{2}}  
\left[ \frac{\im M^{(6)}_{12}}{\Delta m_K } \right]~.
\ee
According to our result in  Eq.~(\ref{eq:chpt_main}), and taking into 
account the estimate of $\xi$ obtained in  \cite{Buras:2008nn}, 
namely $\xi=-(6.0\pm 1.5)\times 10^{-2} \times \sqrt{2} |\epsilon_K|$, the new numerical value of 
 $\kappa_\eps$  is 
\be
\label{eq:keps}
\kappa_\eps = \frac{\sin \phi_\eps}{ 1/\sqrt{2} } \times \left(1 + \rho 
\frac{\xi}{\sqrt{2} |\epsilon_K|}  \right) = 0.94 \pm 0.02~.
\ee
This should be compared with $0.92 \pm 0.02$ in \cite{Buras:2008nn}
and $0.92 \pm 0.01$ in \cite{Laiho:2009eu}, 
where only the long-distance contributions to $\im \Gamma_{12}$ 
(not those to $\im M_{12}$) have been included.

\subsection{Comparison with previous literature}

As anticipated in the introduction, the relative role of short- and long-distance 
contributions to $\epsK$ has been widely discussed in the literature 
in the mid 1980's~\cite{Guberina:1979ix,Wolfenstein:1979wi,Hill:1980ux,Hagelin:1981zk,
Donoghue:1983hj,Hochberg:1983si,Ecker:1984sv,Frere:1985rr,Buras:1985yx}. It is therefore
useful to compare our findings to those in these earlier works.

First of all, we agree on the main conclusion of all these 
papers, namely that $\epsK^{LD}/\epsK^{\rm exp}$ is small as long as $\epsp/\epsK$ is small. 
This is certainly correct, but it is not the point of our analysis: the issue 
we are addressing in this work is the size of the subleading (long-distance) 
contributions to $\epsK$, that vanish in the limit of vanishing $\epsp$.

Second, we agree that single-particle intermediate 
states ($\pi^0$, $\eta$, $\eta^\prime$) do not generate a 
significant long-distance contribution to $\im M_{12}$.
The cancellation of $\pi^0$ and $\eta$ contributions at the lowest order 
in the chiral expansion was noted first in \cite{Donoghue:1983hj}. 
The role of the $\eta^\prime$ was more debated~\cite{Donoghue:1983hj,Hochberg:1983si,
Ecker:1984sv,Frere:1985rr}. The issue was clarified in~\cite{Buras:1985yx}, where it was shown 
that the full nonet contribution  ($\pi^0$, $\eta$, $\eta^\prime$)
vanishes in the large $N_c$ limit. 
This is consistent with our findings, which are based on the 
updated and detailed analysis of the $\eta^\prime$ exchange amplitude 
in Ref.~\cite{Gerard:2005yk}.

Having clarified that single-particle 
intermediate states do not generate a significant contribution
to $\im M^{LD}_{12}$, we are left 
with the two-pion intermediate state as the potentially
leading contribution to $\im M^{LD}_{12}$. 
A naive estimate of this contribution at the partonic level seems to
indicate that it is totally negligible; however, as we have shown, this is
not the case because of the  $\Delta I=1/2$ enhancement of $K \to 2 \pi$ 
amplitudes. Our key observation is that, thanks to chiral symmetry and 
to the $\Delta I=1/2$ dominance, 
the weak phase of this contribution can be related to $\xi$, and 
the problem is shifted to the evaluation of the two-pion
contribution to $\Delta m_K$, as summarised in Eq.~(\ref{eq:main}).
The numerical impact of 
this contribution is then estimated in two ways: i) a direct computation 
of the $\pi\pi$ loop in the leading-log approximation,  
Eq.~(\ref{eq:chpt_main}), 
which provides a {\em definite sign} for this term;
ii) the difference between the experimental value 
of $\Delta m_K$ and the sum of its short-distance contribution
and the other large long-distance contribution provided by the $\eta^\prime$
exchange, which allows us to perform the useful cross-check: 
\be
\D m_K^{LD}|_{G^2_8} \approx  
 \D m_K^{\rm exp} -\left[ \D m_K^{SD} + \D m_K^{LD}|_{\eta^\prime} \right]~.
\ee

We finally note that our estimate of the $\cO(\xi)$ corrections to $\epsK$
is based on the dominance of the $\Delta I=1/2$ amplitude in $K\to 2\pi$
decays. Given the 
experimental smallness of $\Delta I=3/2$ transitions, and the overall size 
of the effect we have evaluated (a few \% correction to $\epsK$),
this is certainly a very safe approximation.

\section{Conclusions} \label{sec:conclusions}
In this paper we have presented a complete analysis of $\epsK$ beyond the 
lowest order in the OPE. In particular, we have analysed the structure of long distance (LD) 
contributions that affect both the {\it absorptive} ($\Gamma_{12}$) and {\it dispersive} ($M_{12}$) 
parts of the $K^0-\bar K^0$ mixing amplitude.
We have pointed out that, in a consistent framework, in addition to 
LD contributions to $\im \Gamma_{12}$, estimated recently in \cite{Buras:2008nn}, also LD 
contributions to $\im M_{12}$ have to be taken into account. Estimating the latter 
contributions in chiral perturbation theory, we found that they reduce by $40\%$ 
the total impact of LD corrections on $\epsilon_K$.

The overall multiplicative factor $\kappa_\epsilon$ in $\epsilon_K$, 
summarising the effect of LD corrections and of the superweak 
phase being different from $45^\circ$, is increased to 
$\kappa_\epsilon=0.94\pm 0.02$, to be compared with $0.92\pm0.02$ obtained without 
LD contributions to $\im M_{12}$. 

\subsection*{Acknowledgments}
We are grateful to Jean-Marc G\'erard and Giancarlo D'Ambrosio for useful comments.
This research was partially supported by the Cluster of Excellence `Origin and Structure 
of the Universe', by the German `Bundesministerium f\"ur Bildung und Forschung' under 
contract 05H09WOE and by the EU under contract MTRN-CT-2006-035482 {\em Flavianet}.

\end{document}